\documentclass[prd,aps,preprint,amsmath,nofootinbib,amssymb,eqsecnum,showkeys,tightenlines]{revtex4-1}
\usepackage{slashed}
\usepackage{epsfig,latexsym,cancel,amssymb,amsmath,verbatim,mathrsfs}
\usepackage{color}
\usepackage{graphicx}

\def\ra{\rightarrow}
\def\L{\left(}
\def\R{\right)}

\def\Ld{\Lambda}
\def\ld{\lambda}
\def\f{\frac}
\newcommand{\be}{\begin{equation}}
\newcommand{\ee}{\end{equation}}
\newcommand{\bea}{\begin{eqnarray}}
\newcommand{\eea}{\end{eqnarray}}
\newcommand{\ba}{\begin{array}}
\newcommand{\ea}{\end{array}}

\long\def\symbolfootnote[#1]#2{\begingroup%
\def\thefootnote{\fnsymbol{footnote}}\footnote[#1]{#2}\endgroup}

\newcommand{\beq}{\begin{equation}}
\newcommand{\eeq}{\end{equation}}

\begin{document}

\title{On UltraViolet Freeze-in Dark Matter during Reheating}
%\title{UV Freeze-in Sterile Neutrino Dark Matter in Scotogenic Models}

\author{Shao-Long Chen}
\email[E-mail: ]{chensl@mail.ccnu.edu.cn}
\affiliation{Key Laboratory of Quark and Lepton Physics (MoE) and Institute of Particle Physics, Central China Normal University, Wuhan 430079, China}

\author{Zhaofeng Kang}
\email[E-mail: ]{zhaofengkang@gmail.com}
\affiliation{School of physics, Huazhong University of Science and Technology, Wuhan 430074, China}

%\\School of Physics, Korea Institute for Advanced Study, Seoul 130-722, Korea}

\date{\today}

\begin{abstract}
The absence of any confirmative signals from extensive DM searching motivates us to go beyond the conventional WIMPs scenario. The feebly interacting massive particles (FIMPs) paradigm provides a good alternative which, despite of its feebly interaction with the thermal particles, still could correctly produce relic abundance without conventional DM signals. The Infrared-FIMP based on the renormalizable operators is usually suffering the very tiny coupling drawback, which can be overcome in the UltraViolet-FIMP scenario based on high dimensional effective operators. However, it is sensitive to the history of the very early Universe. The previous works terminates this sensitivity at the reheating temperature $T_{RH}$. We, motivated by its UV-sensitivity, investigate the effects from the even earlier Universe, reheating era. We find that in the usual case with $T_{RH}\gg m_{\rm DM}$, the production rate during reheating is very small as long as the effective operators dimension $d \leq 8$. Besides, we consider the contribution from the mediator, which may be produced during reheating. Moreover, we study the situation when $T_{RH}$ is even lower than $m_{\rm DM}$ and DM can be directly produced during reheating if its mass does not exceed $T_{MAX}$.
\end{abstract}


\maketitle

\section{Introduction}

Weakly Interacting Massive Particle (WIMP), as the conventional paradigm for dark matter (DM), has been paid extensively attention theoretically and experimentally. However, it faces serious challenge by the null experimental results from sky to underground~\cite{Arcadi:2017kky}. In the WIMP paradigm, DM is thermal produced by the freeze-out mechanism. An alternative paradigm, where the DM never enters the thermal bath and gains its relic density by slowly freeze-in, provides a different solution to the DM puzzle~\cite{{Hall:2009bx, McDonald:2001vt, Bernal:2017kxu}}. Due to its feebly interaction with the standard model (SM) particles, this type DM particles have been named Feebly Interacting Massive Particles (FIMPs). An immediate consequence is that it is of no surprise that the FIMPs are extremely difficult to hunt in the direct detection or collider search experiments. But it may still be detectable in some scenarios, such as the FI$m$P~\cite{Klasen:2013ypa,Kang:2014cia,Queiroz:2014yna,Bae:2017tqn},  a very light FIMP which has a large number density to enhance DM-electron scattering~\cite{Essig:2011nj} or generically decays into photons as a result of without invoking a DM protecting symmetry~\cite{Kang:2014cia}; the keV scale sterile neutrino in the seesaw model is a good case in point for the latter. In addition, there are some special scenarios have been studied where DM could leave cosmological imprints~\cite{Dev:2013yza,Merle:2015oja,Drewes:2015eoa,Heeck:2017xbu,Kainulainen:2016vzv,Roland:2016gli,Tenkanen:2016twd}.

Basically there are two types of FIMP, categorized by the most effective production region. One is called the infra-red (IR-)FIMP DM, whose interactions with the thermal bath are described by the operators ${\cal O}_{d}$ with the operator dimension $d\leq 4$~\cite{Hall:2009bx,Bernal:2017kxu,Chu1,Kaneta:2016vkq,Ayazi:2015jij,Hardy:2017wkr}, and the freeze-in production of DM is most effective at IR (says around the bath particle mass), without be marked much imprints of the very ultraviolet (UV)  Universe. The other one, on the contrary, is characterized by effevtive production at UV by virtue of the DM-bath interactions from high dimensional operators with $d\geq 5$~\cite{Ellis:1984eq,Chung:1998rq,inPQ,Elahi:2014fsa,McDonald:2015ljz,Roland:2014vba,Brdar:2017wgy,Duch:2017khv,Co:2015pka,Garcia:2017tuj}.

The UV freeze-in scenario had not received broad attentions. The reason, probably is ascribed to the need of a clear knowledge of connection to physics up to reheating temperature $T_{RH}$. However, one may think that this offers a way to establish close connections between DM physics and the ultimate early Universe cosmology~\footnote{For instance, it is of importance to build models which could predict reheating temperature and freeze-in FIMP at the same time.}.  Moreover, since the UV freeze-in production mechanism involves high dimensional operators with a high effective cut-off, extremely small coupling is unnecessary. Whereas it is indispensable in the IR freeze-in scenario, which might be considered as its main drawback. Therefore, 
in this sense UV freeze-in is a cure of the issue of IR freeze-in.

In this work, we examine the possible effects from the even early Universe with temperature above $T_{RH}$, namely the reheating era, which did not draw much attention before. Ref.~\cite{McDonald:2015ljz} made first examination based on one particular example and found no significant effects. Here we consider more general cases and find that the effects could be large for some scenarios. It is important to note that the highest
temperature of the early Universe $T_{MAX}$ can be much higher than the reheating temperature $T_{RH}$~\cite{Chung:1998rq}, several effects are 
well expected during that era:
\begin{itemize}
\item Freeze-in production DM during the reheating stage is found to be not significant as long as the effective operator dimension $d\leq 8$;
\item Production of the heavy mediators then translating it number density to DM, which can play a role especially for scenarios with a larger $d$;
\item The production of heavy FIMP DM with mass even far above $T_{RH}$, which is a nontrivial generation of the massive particle production during reheating
to more general cases, with non-renormalizable interactions between DM and the radiation background.
\end{itemize}

After the general cases study we discuss the UV freeze-in effects on concrete models. The tiny neutrino masses are naturally connected to a high seesaw scale.
We present this study on models with DM connecting the origin of neutrino mass. The first example is pretty simple, grounded on operators at the $d=5$ level: The
supersymmetric seesaw portal to the dark sector (minimally a singlet superfield under the dark parity $Z_{2}$) with a seesaw scale $\sim 10^{14}$ GeV. Another example
we study is the well-known scotogenic model~\cite{Ma:2006km}, where the neutrino masses are radiatively generated with the help of TeV scale dark sector
partilces running in the loop. In the model, the DM candidate was normally believed to be the neutral (extra) Higgs doublet component, since 
if the lightest sterile neutrino $N_{1}$ is considered to be DM candidate, a strong tension between the correct relic density with the induced lepton flavor violation 
processes will be drawn. We instead consider a very heavy Higgs doublet mass and choose $N_{1}$ as the dark matter candidate. Such a scenario induces $d=6$ effective
operators describing interacting between the DM and the SM lepton sector after integrating out the heavy Higgs. We find that such a sterile neutrino can be a viable
UV-FIMP DM candidate. The correct relic density can be achieved with a relative low reheating temperature $\sim 10^{3}$ GeV with 
the Higgs doublet mass chosen at $10^{6}$ GeV. We also attempt to obtain the correct DM relic density through the mediator effects or the reheating UV freeze-in.

The paper is organized as the following: In Section II we study the UV freeze-in during reheating stage. In Section III we apply the mechanism to 
DM particle models connecting with the neutrino mass origin. In Section IV we present our conclusions.

\section{UV-FIMP DM production during reheating}

In this section we study UV freeze-in mechanism in different scenarios: (A) High reheating temperature $T_{RH}$ with even higher scale mediators, where the mediators are effectively integrated out; (B) The mediators are around the reheating temperature, and their effects are included; (C) The dark matter mass  is
heavier than the reheating temperature. 

\subsection{Superheavy mediator limit}
In this subsection we consider the case when the portal mediating the dark matter sector and the thermal bath is superheavy. 
We assume the mediator contribution to DM relic density is negligible and consider the effective operators ${\cal O}_d$ by integrating out the heavy 
mediator, where $d$ is the dimension of the operator. For illustration, we will first consider the $d=5$ case then continue to higher dimensional cases.

\subsubsection{The Boltzeman equations (BEs)}
The UV freeze-in mechanism is sensitive to the very early Universe. However, the hottest phase, the reheating phase was used to be neglected. 
DM is assumed to be produced within the radiation-dominated era below the reheating temperature $T_{RH}$,
\begin{align}\label{}
T_{RH}=2.4\times10^{5}\times\L\f{100}{g_*}\R^{1/4}\L\f{M_\phi}{10^8\rm GeV}\R^{1/2}\L\f{\alpha_\phi}{10^{-14}}\R^{1/2}\rm GeV,
\end{align}
where $g_{*}$ is the effective number of relativistic degrees of freedom at this era, $M_{\phi}$ and $\Gamma_{\phi}\equiv \alpha_{\phi} M_{\phi}$
are the mass and decay width of the inflaton respectively. One should wonder if this is an acceptable approximation. The $d=5$ example studied in 
Ref.~\cite{McDonald:2015ljz} suggests that is is indeed acceptable. We will show that this is also true for the larger $d$ as long as $d\leq 8$, which
indicates that in such scenarios DM freeze-in production during reheating are dominated by the IR region.

To have a closer look, let us start from the BEs of the reheating dynamics which is described by the following  two coupled equations of energy
densities~\cite{Chung:1998rq}:
\begin{align}\label{BE:early1}
&\dot{\rho}_\phi+3H\rho_\phi+\Gamma_\phi \rho_\phi=0,\\
&\dot{\rho}_R+4H\rho_R -\Gamma_\phi \rho_\phi=0,\label{BE:early2}
\end{align}
where $\phi$ and $R$ denote the inflaton (non-relativistic matter) and SM radiation, respectively.  As usual we assume that all radiation components respect the thermal distribution,~\footnote{The SM fermions may be not direct product of inflaton decay, but they were assumed to be thermalized immediately via scattering or decay.}
\begin{align}\label{}
f^{eq}(E,T)=\f{1}{e^{E/T}\pm1},
\end{align}
with $``-"$ and $``+"$ for the fermion and boson respectively, an approximation $f^{eq}(E,T)=e^{-E/T}$  is taken in this work.

The evolution of the system can be simplified using the approximate picture of ``inflaton dominance" during the preheating era, namely when $H(a)>\Gamma_\phi$. In this picture, inflaton energy overwhelmingly dominated over others and inflaton decay products (typically the SM particles) formed the radiation thermal bath, whose backreaction effects on inflaton are not took into account. Such a treatment facilitates decoupling between two equations. Thereby, Eq.~(\ref{BE:early1}) is trivially solved by $\rho_\phi\propto a^{-3}$ (purely redshift), which is putative to work until $H(a_{RH})\sim\Gamma_\phi$. While the thermal bath had been receiving the radiation energy from inflaton decay gradually. By approximately solving Eq.~(\ref{BE:early2}), one obtains the scaling behavior of radiation energy density $\rho_R(T)=\f{\pi^2}{30}g_*T^4$ with the scale factor; concretely, it leads to the following behavior of $T$ with the scale factor~\cite{Chung:1998rq}:
\begin{align}\label{preH}
\f{T}{M_\phi}\simeq c_{1}^{1/4}\L\frac{12}{\pi^{2}g_{*}}\R^{1/4} \L \frac{3}{8\pi}
\frac{M_{Pl}^{2}H_{I}^{2}}{M_{\phi}^{4}}\R^{1/8}   \left[ (a/a_I)^{-3/2}-(a/a_I)^{-4}\right]^{1/4} \quad (H\gg \Gamma_\phi),
\end{align}
with $a_I$ the initial size of the Universe and $c_{1}=\sqrt{3/8\pi}\alpha_{\phi}(M_{Pl}/M_{\phi})$, with the Planck mass 
$M_{Pl}=1.22\times 10^{19}$ GeV . 
The initial Hubble parameter $H_{I}$ depends on the inflation model and it is chosen to be around $M_\phi$ in the simplest chaotic inflation model.
It can be seen that the highest temperature was reached as the size $a=1.48 a_I$,
\begin{align}\label{}
\f{T_{MAX}}{T_{RH}}=0.77\L\f{9}{5\pi^3g_*}\R^{1/8}\L\f{H_IM_{Pl}}{T_{RH}^2}\R^{1/4}\,.
\end{align}
After the Universe temperature reached $T_{MAX}$, it began to decrease, following the power law $T\propto a^{-3/8}$ until $T_{RH}$ and the evolution
of the energy density was $\rho_{\phi}\propto a^{-3}\propto T^{8}$ and consequently $H(T)\propto T^{4}$; more concretely, $H=C_{p}T^{4}$ with
\begin{align}\label{}
C_p=6.88g_{*} M_{\phi}^{-1} M_{Pl}^{-2}\alpha^{-1}_\phi\,.
\end{align}
We noticed that this is about 5 times smaller than the one used in Ref.~\cite{McDonald:2015ljz}. It can be rewritten as 
$C_{p}=20.8 \sqrt{g_{*}}/(M_{Pl}T_{RH}^{2}) $, to demonstrate its dependence on $T_{RH}$.
 
Now we study the BE for DM, which is denoted as $X$. The DM number density $n_{X}$ keeps accumulating after inflation via the thermal scattering or decay.
For simplicity, the relevant processes is this study are assumed to be four-body, says $A+B\ra X+X^*$ or $A\ra B+X+X^*$,~\footnote{In this treatment we 
neglect the back reaction to the inflaton-radiation system. This is reasonable since DM took up a tiny energy fraction during reheating.} 
with $A$ and $B$ are thermal particles; variants like $A+B\ra X+C$, etc., are possible in different contexts. Here we focus on the first example and the 
corresponding BE is given by
\begin{align}\label{BE:N}
&\dot{n}_X+3Hn_X= \int d\Pi_A d\Pi_{B}d\Pi_Xd\Pi_{X^*}(2\pi)^4\delta^{(4)}(p_X+p_{X^*}-p_A-p_{B})|{\cal M}|^2 f_A^{eq}f_{B}^{eq},
\end{align}
where $d\Pi={g}/(2\pi)^3\cdot d^3p/2E$ with $g$ the internal degrees of freedom for a given species. 
The right-handed side contains the collision terms that could create or destroy $X$, which the latter, i.e., the process 
$ X X^*\ra AB $ can be safely neglected. 
The right-handed side of the above equation is known as the interaction density $\gamma(T)$ that will be specified later. 
 
As usual, the Hubble expanding effect can be absorbed by considering the dimensionless quantity, $X_{X}\equiv n_{X}a^{3}$, the two terms in the left-handed side of Eq.~(\ref{BE:N}) can be combined into a single term $(H/a^{2}) dX_{X}/da$. Consequently, the BE can be rewritten as
\begin{align}
\frac{dX_{X}}{dT}=\frac{a^{2}}{H}\frac{da}{dT}\gamma(T)=\frac{1}{3H}\frac{da^{3}}{dT}\gamma(T)\,,
\end{align}
where the differential variable $a$ has been traded with temperature which has more transparent physical meaning in our context. This trading should also be
implemented on other quantities of function of  $T$ (such as $H$ and $a$), utilizing the following relations in reheating (RH) and radiation dominated (RD) era
\begin{align}
\text{RH :} &\quad \frac{2}{3t}=H=C_{p}T^{4}\propto a^{-3/2} \quad\text{with}\quad C_{p}=6.88 g_{*}M_{\phi}^{-1} M_{Pl}^{-2}\alpha_{\phi}^{-1}\,,
\nonumber \\
\text{RD :} & \quad \frac{1}{2t}=H=C_{r}T^{2}\propto a^{-2} \quad\text{with} \quad C_{r}=1.66 \sqrt{g_{*}}M_{Pl}^{-1}\,.
\end{align} 
The BEs in the two eras can be rewritten respectively as 
\begin{align}\label{BE:11}
\frac{d(n_{X} T^{-3})}{dT}=-\frac{C_{r}^{-1}}{T^{6}} \gamma(T) \quad\text{and}\quad
\frac{d(n_{X} T^{-8})}{dT}=-\frac{8}{3}\frac{C_{p}^{-1}}{T^{13}}\gamma(T)\,
\end{align}
where the first equation is nothing the evolution of $n_{X}/s$  (with $s$ the entropy density), the widely used yield of a species. It keeps constant for 
$\gamma \to 0$. Whereas during RH phase it is $n_{X} T^{8}$ that keeps constant for $\gamma \to 0$, attributed to entropy production from inflation decay,
which results in $a^{3}\propto T^{-8}$ instead of  $a^{3} \propto T^{-3}$ as in the RD era.

\subsubsection{Case $d=5$}

%The UV freeze-in mechanism is sensitive to cosmology in the UV era, so reheating after inflation may produce negligible effects on the FIMP production. To investigate that, we consider a system including the classical inflaton filed $\phi$ with decay width $\Gamma_\phi$ into the SM radiation $R$ but not to the new fields $\Phi$ and DM. After summation over the spins of final states,
Before specifying the concrete form of $\gamma(T)$, it is helpful to analyze some general features of DM productions in two eras based 
on the BEs in Eq.~(\ref{BE:11}). Integrating over $T$ until $T_{MAX}$, one obtains
\begin{equation}\label{d5:1}
\frac{n_{X}}{T^{3}}=\int^{T_{RH}}_{T}\frac{1}{C_{r}} \frac{\gamma(T)}{T^{6}} dT+
\f{8}{3}\frac{T^{5}_{RH}}{C_{p}}\int^{T_{MAX}}_{T_{RH}} \frac{\gamma(T)}{T^{13}} dT\,.
\end{equation}
It clearly tells us that the first term is UV-sensitive provided that $\gamma(T) \sim T^{n}$ with $n\geq 6$, which corresponds to $d\geq5$ based on dimension
analysis. By contrast, the second term, the production yield during reheating, is IR-sensitive as long as $d\leq 8$, which we will show later. The reason is traced 
back to the much slower temperature dropping during the RH era, $T\propto a^{-3/8}$ , compared to the followed era, $T\propto a^{-1}$. From Eq.~(\ref{d5:1})
and utilizing entropy conservation, one can get $\f{n_{X}(T_{0})}{T_{0}^{3}}=\f{n_{X}(T)}{T^{3}}\f{T^{3}a(T)^{3}}{T_{0}^{3}a_{0}^{3}}=\f{g_{*}(T_{0})}{g_{*}(T)}\f{n_{X}(T)}{T^{3}}$ 
with $g_{*}(T_{0})\approx 3.9$ and $T_{0}=2.37\times 10^{-13}$ GeV, where $T$ is any temperature substantially below $T_{RH}$ but far above the weak 
scale ( thus $g_{*}(T) = g_{*}(T_{RH}) $). Its concrete value is irrelevant due to IR-insensitive property of the first integral in Eq.~(\ref{d5:1}). Then the final
relic density of DM with mass $M_{X}$ is expressed as 
\begin{equation}\label{d5:omega}
\Omega_{X} h^{2}=\f{M_{X} n_{X}(T_{0})}{\rho_{c}} h^{2} =\f{g_{*}(T_{0})}{g_{*}(T_{RH})} \frac{M_{X}T_{0}^{3}}{\rho_{c}}
\left[ \frac{n_{X}(T)}{T^{3}}\right]_{T\lesssim T_{RH}} h^{2}\,,
\end{equation}
where the critical energy density today is $\rho_{c}=8.1\times 10^{-47} h^{2}\,\text{GeV}^{4}$.

As the leading effective operator for freeze-in production, we consider the $d=5$ case.
They could be $\f{1}{\Ld}|H|^2\bar\chi\chi$ or $\f{1}{\Ld}\bar f f |S|^2$ for a fermionic or scalar FIMP, respectively. 
Besides, operators involving DM derivative couplings, such as gravitino and axion-like particles, are also well-known examples~\cite{Ellis:1984eq, inPQ}.
Barring exceptional examples like resonance enhancement and so on, without knowing the details of the operators  the $S$-matrix of the scattering process
is supposed to take the form of
\begin{align}\label{}
|{\cal M}|^2\sim \f{s}{\Ld^2}.
\end{align}
The scaling behavior of amplitude with $s$ is kept explicit while the other model dependent factors have been absorbed  into the redefinition of  $\Ld$. 
This expression usually is good for massless final states, it maybe be modified in a more model dependent way, and we list some examples in the Appendix. 

Now with the amplitude squared we are able to calculate $\gamma(T)$.
For the $2\ra 2$ process $A+B\ra X + X^{*}$, one of  the standard methods~\cite{Davidson:2008bu} to deal with $\gamma(T$ is 
separating the center of mass from initial state phase space integrals by inserting identity 
$1=\int d^4Q \delta^4(Q-p_A-p_{B})=\f{1}{2}\int\sqrt{Q_0^2-s}ds d\Omega dQ^0\delta^4(Q-p_A-p_B)$ and 
further using  the two-body phase space
\begin{align}\label{}
\int  d\Pi_A d\Pi_B (2\pi)^4\delta^4(Q-p_A-p_B) =\f{\sqrt{(p_A\cdot  p_B)^2-m_A^2m_B^2}}{4\pi \sqrt{Q^2}}\approx \f{1}{8\pi},
\end{align}
where the last approximation holds in the massless limit, which is always good in the UV freeze-in scenario. Moreover, the two-body phase space of the final states in the center of mass frame is given by
\begin{align}\label{}
\int  d\Pi_X d\Pi_{X^{*}} (2\pi)^4\delta^4(Q-p_X-p_{X^{*}}) =\int d\Omega_X \f{1}{16\pi^2}\f{2|\vec p_X|}{\sqrt{Q^2}}
\approx  \int d\Omega_N\f{1}{16\pi^2}.
\end{align}
Then the reaction density can be written as the single integral over $s$:
\begin{align}\label{}
\gamma\approx &\f{g_A g_B}{(2\pi)^4} \f{1}{8\pi\times 16\pi^2}\int \f{1}{2}ds d\Omega \int d\Omega_X  |{\cal M}|^2 \int^\infty_{\sqrt{s}} dQ^0 e^{-Q_0/T} \sqrt{Q_0^2-s}\cr
=&\f{T}{\Ld^2} \f{g_Ag_B}{1024\pi^5}\int_0^\infty ds  s^{3/2}K_1(\sqrt{s}/T).
\end{align}
The integral can be done analytically, giving $ 2\times 2^4\times\Gamma[3]\Gamma[2]  T^5= 32 T^5$, and thus eventually the rate is calculated to be
\begin{align}\label{}
\gamma=\f{g_A g_B}{32\pi^5} \f{T^6}{\Ld^2}\,.
\end{align}

Now substituting it into Eq.~(\ref{d5:1}) then Eq.~(\ref{d5:omega}), we obtain the following expression for the DM relic density
\begin{align}\label{d5:relic}
\Omega_{X} h^{2} & = \frac{g_{*}(T_{0})}{g_{*}(T)} \frac{M_{X} T_{0}^{3}}{\rho_{c}} \frac{g_{A}g_{B}}{32\pi^{5}}
\frac{T_{RH}}{C_{r} \Lambda^{2}} (1+\delta_{re}) h^{2} \cr
&= 0.116\times \L\f{g_{A}g_{B}}{4}\R \L\f{M_{X}}{100\text{GeV}}\R 
\L\f{T_{RH}}{10^{8} \text{GeV}}\R \L\frac{10^{16} \text{GeV}}{\Lambda}\R^{2}\,.
\end{align}
It indicates that typically a very high (effective) suppression scale $\Lambda \gg 10^{10}\text{GeV}$ is required to suppress the weak scale DM yield.
In the above estimate we have used the fact that the relative size of the reheating phase contribution is very small:
\be
\delta_{re}=\frac{4}{9} \frac{C_{r}}{C_{p}}\frac{1}{T_{RH}^{2}}\approx 0.036 \ll 1\,.
\ee
It is roughly a constant, provided that the simple estimate on $T_{RH}$ is not modified. This result matches the one calculated in 
Ref.~\cite{McDonald:2015ljz} despite of some different intermediate steps.

\subsubsection{Beyond $d=5$}
The previous discussions can be easily generalized to the cases with higher dimension $d>5$ operators ${\cal O}_{d}$ and
we will see that the DM freeze-in production during the reheating phase does not have a substantial contribution as long as $d\leq 8$.
Let us assume that the amplitude squared takes the general form $|{\cal M}|_{d}\sim (s/\Lambda^{2})^{d-4}$. Then with assuming the reactions are through 
$2\to 2$ processes,  the reaction density can be calculated to be~\footnote{Actually the reactions may involve final states with more than two particles, which
is particularly true for larger $d$. But that just complicates the phase space integral and gives extra numerical factors, which again can be absorbed into the redefinition of  $\Lambda$, see one demonstration in Ref.~\cite{Elahi:2014fsa}.}
\begin{align}\label{beyond5:1}
\gamma=\f{T}{\Ld^{2(d-4)}} \f{g_A g_B}{1024\pi^5}\int_0^\infty ds  s^{(2d-7)/2}K_1(\sqrt{s}/T)
=g_A g_B\f{(d-3)!(d-4)!}{2^{15-2d}\pi^5} \f{T^{2d-4}}{\Ld^{2d-8}}\,.
\end{align}
Increasing $d$ leads to the stronger dependence on temperature of reaction rate, thus more sensitivity to UV scale.
But as long as $d< 8$, the integral over $T$ in the reheating phase is always dominated by IR. As a consequence, the DM relic density is given by
\be
\Omega_{X} h^{2} = g_{A} g_{B} \frac{g_{*}(T_{0})}{g_{*}(T_{RH})} \frac{M_{X}T_{0}^{3}}{\rho_{c}}
\f{(d-3)!(d-4)!}{(2d-9)2^{15-2d}\pi^5} \frac{T_{RH}^{2d-9}}{C_{r} \Lambda^{2d-8}}(1+\delta_{re}) h^{2}\,,
\ee
where now the ratio between the contributions during RD and RH eras becomes
\be
\delta_{re}=-\frac{8}{3} \frac{2d-9}{2d-16}\frac{C_{r}}{C_{p}}\frac{1}{T_{RH}^{2}}\,,
\ee
which is valid for $d<8$. Increasing $d$ leads to a mildly larger $\delta_{re}$, but the relative size is still set by the factor $C_{r}/(C_{p}T_{RH}^{2})$,
a small value as in the $d=5$ case.~\footnote{Similar results were obtained before in Ref.~\cite{Co:2015pka,Garcia:2017tuj}, and we thank the authors to inform us that. }

The case with $d=8$ makes a real difference. Now the integration in the range $[T_{RH}, T_{MAX}]$ region leads to an enhancement factor
$\log(T_{MAX}/T_{RH})$, developing UV sensitivity on $T_{MAX}$. But the logarithmic enhancement ($\sim {\cal O}(10)$) does not overcome
the small factor $C_{r}/(C_{p}T_{RH}^{2})$ and thus the freeze-in production is still dominated by RD phase. Whereas for $d>8$ case, DM production during 
reheating tends to dominated over that during radiation-dominated era
\be
\delta_{re}=\frac{8}{3} \frac{2d-9}{2d-16} \L\frac{C_{r}}{C_{p}}\f{1}{T_{RH}^{2}}\R 
\L\frac{T_{MAX}}{T_{RH}}\R^{2d-16}\,,
\ee
where the enhancement factor can be large as $(T_{MAX}/T_{RH})^{2}\sim 10^{6}$ for $d=9$. Of course, whether such high dimension operator is
of practical interests is another issue. In this work we just point out that the UV freeze-in mechanism with a fairly large $d$
could works more effectively during the RH era rather than the conventional RD era.

\subsection{Producing the mediator during reheating}

The effective operators ${\cal O}_d$ will lose its feasibility in the very high temperature region, since the mediator $\Omega$,  which is heavier than $T_{RH}$ 
and has been integrated out, turns out to be active. Moreover,  when the mediator $\Omega$ is a member of the dark sector and its main decay 
channel is into the FIMP DM, one has to consider the roles played by the mediators in determining the DM final relic density. 

We consider the scenario with the mediator mass $T_{MAX}\geq M_\Omega\gg T_{RH}$. It can be even heavier than $T_{MAX}$, 
however, that case renders the yield of $\Omega$ suppressed and thus is of no interest here. 
The thermal radiation produces $\Omega$ via scatterings, says $f\bar f\ra \Omega\Omega^*$ if it is charged under the SM gauge groups. 
The thermal average of the scattering cross section times relative velocity is parameterized as $\langle \sigma v\rangle =\alpha_\Omega /M^2_\Omega$. 
Then, the BE for energy density, under the decoupling approximation, is simply given by
\begin{align}\label{}
&\dot{\rho}_\Omega+3H\rho_\Omega+ \f{\langle \sigma v\rangle_\Omega }{m_\Omega} \left( \rho_\Omega^2-(\rho^{eq}_\Omega)^2 \right) -\Gamma_\Omega (\rho_\Omega-\rho_\Omega^{eq})=0.
\end{align}
The scenario we consider here share many features as the one in the Ref.~\cite{Chung:1998rq}. The main difference is that here the superheavy 
particle $\Omega$ is chosen to be a decaying particle. Let us assume that it dominantly decays into the FIMP DM, with a decay width 
$\Gamma_\Omega$, which could be larger or smaller than $1/H(T_{RH})$. But in practice the decay of $\Omega$ does not matter because again $\Omega$ just 
took up a tiny energy fraction in the preheating era and it never came to chemical equilibrium with the radiation thus having insignificant impact on others except 
for the DM number density. If $\Gamma_\Omega\ll 1/H(T_{F})$ with $T_F$ the freeze-out temperature of $\Omega$ which was much above $T_{RH}$, the 
losing (decay) term can be removed and then it is exactly reduced to the one in Ref.~\cite{Chung:1998rq},
the total number density of $\Omega$ at $T_{F}$ is given by
\be
n_{\Omega}(x_{F}) a_{F}^{3} = 2.6\times 10^{-3} a_{I}^{3} M_{\phi}^{3}\frac{\alpha_{\phi}^{3} \alpha_{\Omega}}{g_{*}^{3}}
\L\frac{H_{I}^{2}M_{Pl}^{6}}{M_{\Omega}^{8}}\R \equiv a_{I}^{3} A^{3}\,,
\ee
which is just suppressed by powers of $T_{RH}/M_{\Omega}$ instead of exponentially, as a result of the largeness of $\lambda\gg 2/17$, with
\be
\lambda \equiv \frac{2^{5/4}\pi^{3/4}g_{*}^{1/4}}{\sqrt{3}} \L\frac{M^{4}_{\Omega}}{\alpha_{\phi} M_{\phi} H_{I} M_{Pl}^{2}}\R^{1/4}\,.
\ee
It determines the freeze-out scale factor $a_{F}/a_{I}=(17/2\lambda)^{8/3}$.

Using the fact that $n_{F}a_{F}^{3}$, the total number of $\Omega$, is a comoving constant, its contribution to DM energy density today is 
\be
\Delta\rho_{\Omega}(T_{0})=M_{\Omega}n_{\Omega}(T_{0})=M_{\Omega}A^{3}\L\frac{a_{I}^{3}}{a_{RH}^{3}}\R 
\L\frac{a_{RH}^{3}}{a_{0}^{3}}\R\,.
\ee
Since in the RH and RD eras the Universe followed different evolution rules, we accordingly split the ratio $a_{I}^{3}/a_{0}^{3}$
into two parts. The ratio in the first bracket can be calculated in terms of  Eq.~(\ref{preH}),
\be
\frac{a_{I}^{3}}{a_{RH}^{3}}\approx \L\frac{12}{\pi^{2}g_{*}}\R^{-2} c_{1}^{-2} 
\L\frac{3}{8\pi} \frac{M_{Pl}^{2} H_{I}^{2}}{M_{\phi}^{4}} \R^{-1}\L\frac{T_{RH}}{M_{\phi}}\R^{8}
=\frac{400\pi^{2}M_{\phi}^{2}}{H_{I}^{2}}\alpha_{\phi}^{2}\,.
\ee
While the second ratio can be obtained by taking the comoving entropy conservation after reheating, $a_{RH}^{3}/a_{0}^{3}=T_{0}^{3}g_{*}(T_{0})
/g_{*}(T_{RH})T^{3}_{RH}$. Combining all these factors together, one arrives the corresponding contribution to the DM fraction,
\be
\Delta\Omega_{X}h^{2}=\frac{\Delta \rho_{X}(T_{0})}{\rho_{c}}h^{2}=0.03 \L\frac{M_{X}}{1\text{TeV}}\R
\L\frac{10^{2.5}T_{RH}}{M_{\Omega}}\R^{8} \L\frac{10^{6} \text{GeV}}{T_{RH}}\R \L\frac{\alpha_{\Omega}}{0.01}\R
\L\frac{106}{g_{*}}\R^{3/2}\,.
\ee
For the above parametrization one has 
\be
\f{T_{MAX}}{T_{RH}}\approx 1.8\times 10^{3}\L \f{m_\phi}{10^8\rm GeV}\R^{1/4},\quad 
\alpha_\phi=2.8\times10^{-15}\L\f{ T_{RH}}{10^6\rm GeV}\R^2 \L\f{10^8\rm GeV}{m_\phi}\R.
\ee
As expected, the relic density is very sensitive to $M_{\Omega}$, and it would blow up as $M_{\Omega}$ is just near $T_{RH}$.

\subsection{Heavy dark matter confronting low reheating temperature}
In this subsection we move to another UV freeze-in scenario: DM itself is heavier than the reheating temperature $T_{RH}$ and consequently its 
yield during the RD era is highly suppressed; on the other hand, it is still lighter than $T_{MAX}$ and thus it can be produced during RH phase. This nothing but 
a generalization of massive particle production during reheating with constant $\langle\sigma |v|\rangle$, studied in the previous subsection, to the case with
a nontrivial $\langle\sigma |v|\rangle$. Specifically, we focus on the case that ${\cal O}_{d}$ is still a valid description of DM interactions during RH
and hence the reaction density $\gamma$ can be read from Eq.(~\ref{beyond5:1}). Of course, it is reliable only in the relative high temperature region
$T> M_{X}$. A more precise calculation should take into account the heavy DM mass threshold, which will be done in what follows.

Let us recalculate the reaction density for a heavy $X$ produced via the collision $A+B\ra X+X^{*}$, fully taking into account the threshold effect. The
amplitude squared is assumed to be generic $|{\cal M}|_{d}^{2}=(s/\Lambda^{2})^{d-4}$. For fermionic DM, $s^{d-4}$ may be replaced by 
$(s-4M_{X}^{2}) s^{d-5}$ depending on the structure of the fermionic bilinear operator. Here we focus on the simplest case and put others in the Appendix.
The procedure is the same with the massless case, except that the final two-body phase space should be modified as
\be
\int d\Pi_{X} d\Pi_{X^{*}}(2\pi)^{4}\delta^{4}(Q-p_{X}-p_{X^{*}})=\int d\Omega_{X} \frac{1}{16\pi^{2}} \sqrt{1-4M_{X}^{2}/s}\,.
\ee
The new factor means that only the collision with center-of-mass energy $\sqrt{s} > 2 M_{X}$ is allowed. The last step is calculating the integration
\be
\gamma=\f{T}{\Ld^{2(d-4)}} \f{g_A g_B}{1024\pi^5}\int_{2M_{X}}^\infty ds  s^{(2d-7)/2}\sqrt{1-4M_{X}^{2}/s}K_1(\sqrt{s}/T)\,,
\ee
which admits an analytical expression in terms of the MeijerG function (We choose the software Mathematica for this job.)
\be
\gamma=\frac{T^{2d-4}}{\Lambda^{2(d-4)}} \frac{g_{A}g_{B}}{1024\pi^{5}}\frac{\sqrt{\pi}}{2} \L\frac{2M_{X}}{T}\R^{2d-6}
\text{MeijerG}[\{\{\},\{\frac{9}{2}-d\},\{\{0,1,3-d\},\{\}\},\frac{M_{X}^{2}}{T^{2}}]\,.
\ee
The MeijerG function has tow analytical limits when $M_{X}/T$ taking very small and extremely large values. For small $M_{X}/T$, it is verified that
it gives the same result given in Eq.~(\ref{beyond5:1}). For large $M_{X}/T$, the function can be expanded as
\be\label{meijerG}
\text{MeijerG}[\{\{\},\{\frac{9}{2}-d\},\{\{0,1,3-d\},\{\}\},\frac{z^{2}}{4}]=e^{-z/2}\frac{\sqrt{\pi}}{z} \left[ 2 +\frac{3(2d-7)}{z}+{\cal O}(1/z^{2}) \right]\,.
\ee
What is more, we find that if the expansion is terminated at $(1/z)^{2d-7}$, the approximation also works in the $2M_{X}/T \gg 1$ region, with error within 
a few percents.
Thus a larger $d$ requires more terms for the sake of a sufficiently good approximation. We cast the remaining terms in the appendix for several typical cases.

Let us analyze the integral over $T$ given in Eq.~(\ref{d5:1}). The first term can be neglected if $T_{RH}\ll M_{X}$, owing to the exponential suppression 
factor $e^{-M_{X/T}}$ in Eq.~ (\ref{meijerG}). Whereas production during RH is not suppressed if $M_{X} < T_{MAX}$. In the massless DM limit, 
DM freeze-in during RH is not sensitive to UV (for $d < 8$), as is not changed in the presence of a heavy DM mass. This is easy to understand, in higher 
temperature region $T_{MAX} > T\gg M_{X}$, the reaction density effectively tracks the $T$-dependence in the massless limit. 
The final result is explicitly given by
\be
\frac{n_{X}}{T^{3}}|_{T=T_{RH}}\approx \frac{8}{3}\frac{T^{5}_{RH}}{C_{p}}\int_{T_{RH}}^{T_{MAX}}\frac{\gamma(T)}{T^{13}}
\approx 0.03\times \frac{T^{5}_{RH}}{M_{X}^{6}\Lambda^{2}C_{p}}\,,\qquad 0.2\times \frac{T_{RH}^{5}}{M_{X}^{4}\Lambda^{4}C_{p}}\,,
\ee
for $d=5$ and $d=6$ cases respectively. As expected, the integral is insensitive to neither $T_{RH}$ nor $T_{MAX}$, provided that $M_{X}$ is sufficiently lighter than $T_{MAX}$. Otherwise a substantial correction is expected for $M_{X}$ near $T_{MAX}$. We can estimate the typical parameters for the correct
relic density
\be
\Omega_{X}h^{2}=\frac{g_{*}(T_{0})}{g_{*}(T_{RH})} \frac{M_{X}T_{0}^{3}}{\rho_{c}}h^{2}
\L 0.03\frac{T^{5}_{RH}}{M_{X}^{6}\Lambda^{2}C_{p}}\,,\;\; 0.2 \frac{T_{RH}^{5}}{M_{X}^{4}\Lambda^{4}C_{p}}\R\,.
\ee
If $M_{X}$ is not heavier than $T_{RH}$ in many orders of magnitude, a high scale $\Lambda$ is needed to suppress $\Omega_{X}h^{2}$.

\section{Close connections to neutrino physics}
As applications of the above general analysis, in this section we present two concrete UV-FIMP dark matter candidates.

\subsection{Supersymmetric high scale seesaw portal}
The leading order operator indicates a very high cutoff scale near the GUT scale, which brings a hint
for FIMP DM connected with the high scale seesaw portal. The high scale seesaw mechanism naturally explains the tiny
neutrino mass. The most natural realization of such type of UV-FIMP DM is the supersymmetric seesaw mechanism, extended with
a singlet superfield $S$ which is odd under dark parity $Z_{2}$~\footnote{Even earlier model building for IR-FIMP by extending the supersymmetric
seesaw model was given in Ref.~\cite{FIMPdecay}.}. The most general superpotential reads
\be
{\cal W}={\cal W}_{\rm MSSM} + \frac{1}{2}\lambda_{sn}S^{2}N + y_{N}LH_{u}N + \frac{M_{N}}{2}N^{2} +\frac{M_{S}}{2}S^{2}\,, 
\ee
where $M_{S}$ is at the weak scale while $M_{N}\leq 10^{14}$ GeV. Any component of  $S=(\widetilde{S}, R_{s}, A_{s})$ can be the lightest one, depending on the soft supersymmetry breaking parameters. We assume that the Majorana fermion $\widetilde{S}$  is the lightest one. The other components could also contribute to the DM relic density, as will be shown in the following.

Integrating out the heavy $N$ filed, the effective low energy theory is described by the MSSM$ + S$ plus three types of dimension-five operators
\be
{\cal W}={\cal W}_{\rm MSSM} +\frac{M_{S}}{2}S^{2} + \frac{y_{N}^{2}(LH_{u})^{2}}{M_{N}}+y_{N}\lambda_{sn}\frac{S^{2} LH_{u}}{M_{N}}
+\lambda_{sn}^{2}\frac{S^{4}}{M_{N}}\,.
\ee
Among these dimension-five operators, the first provides the tiny neutrino mass and mixings, the second is accounting for the DM relic density via UV freeze-in, through processes such as the scattering $\widetilde{L} + H_{u} \ra \widetilde{S} + \widetilde{S}$. The DM relic density is given by in Eq.~(\ref{d5:relic}). One may note that the $S^{2}LH_{u}$ term also furnishes decay channels for $R_{s}$ and $A_{s}$, for instance the three-body decay pattern $R_{s}\ra \widetilde{S}+\ell +H_{u} $, and eventually contribute to the total DM relic abundance. The usual $R$-parity is violated, 
but the effect is suppressed by the heavy RHN. 

\subsection{d=6: The scotogenic model with a heavy Higgs doublet}
The second example is based on the $d=6$ operator $\frac{1}{\Lambda^{2}} \bar f f \bar XX$ for a fermionic DM interacting with the SM fermions, obtained by integrating out the heavy scalar mediators, which provide the bridge between the SM and dark sector. We focus on an explicit model called the ``scotogenic'' model for radiative neutrino mass generation~\cite{Ma:2006km}.

\subsubsection{A quick review of the model}
The model introduces a second Higgs doublet $\Phi$ and three families of right-handed neutrinos $N_i$ ($i=1, 2, 3$), which all transform odd under discrete symmetry $Z_2$. In the basis where the three generations of RHN masses are diagonal, the most general Lagrangian is written as
\begin{align}\label{}
{\cal L}={\cal L}_{SM} + \L Y_{\alpha i}^N\bar \ell_\alpha \Phi N_i+\f{M_i}{2}\bar N_i N_i^C+h.c.\R+V_{SM}(\Phi, H),
\end{align}
with $M_1<M_2<M_3$. The Higgs potential $V(\Phi, H)$ contains the following terms
\begin{align}\label{}
V(\Phi, H)=m_\Phi^2|\Phi|^2+\f{\ld_5}{2}\left[(\Phi^\dagger H)^2+h.c.\right]+V(H),
\end{align}
where $H$ is  the SM Higgs doublet.
For the limit $m_\Phi\gg {\rm TeV}$,  the quartic couplings, some of which can induce mass splittings between the components of $\Phi$,
become irrelevant. The $\ld_5$-term is required for neutrino mass generation. In the limit $m_\Phi\gg M_i$, the resulting neutrino mass matrix 
is given by~\cite{Ma:2006km}
\begin{align}\label{nu:mass}
(M_\nu)_{\alpha\beta}\approx \f{\ld_5 v^2}{16\pi^2 m_\Phi^2}\sum_kY_{\alpha k}^NY_{\beta k}^N M_k
\end{align}
The add up of $\Phi$ and $N_{i}$ will introduce lepton flavor violation (LFV) processes at loop level. The LFV rates 
is stringently constrained by the current searches. For instance, the strictest bound set for $\mu\to e\gamma$ is 
${\rm Br}(\mu\to e\gamma) \lesssim 5.7\times 10^{-13}$~\cite{Adam:2013mnn}.

In the minimal model, this branching ratio is calculated to be~\cite{Kubo:2006yx},
\begin{align}\label{}
{\rm Br}(\mu\ra e\gamma)&=\f{3\alpha}{64\pi (G_F m_\Phi^2)^2}|Y_{\mu k}^NY_{ek}^{N*} F_2(M_k^2/m_\Phi^2)|^2,\cr
F_2(x)&=\f{1-6x+3x^2-6x^2\ln x}{6(1- x)^4},
\end{align}
with $F_2(x)\ra 1/6$ in the limit of $M_k^2\ll m_\Phi^2$. It is easy to see that the heavy $\Phi$ greatly suppresses the branching ratio and therefore it is easily made below the upper bound for $m_\Phi\sim \rm PeV$ even for $Y^N\sim {\cal O}(1)$.

\subsubsection{UV-FIMP with low reheating temperature scenario}

In this model the sterile neutrino $N_i$ is difficult to be a good thermal WIMP DM candidate. The reason is ready to understand. On the one hand, $N_i$ interacts with the SM particles only through $\Phi$-mediation, with strength $Y^N_{\alpha_i}$, which, to achieve DM correct relic density after freezing-out, is supposed to be order one for a weak scale. On the other hand, that light while large $Y^N_{\alpha_i}$, generically renders the LFV rates exceeding the upper bound (See a rescue in Ref. [17].). Therefore, usually the lighter neutral component of the inert Higgs doublet is regarded as the DM candidate in this model. However, if we give up the requirement that DM should be thermal, the sterile neutrinos provide a good example of UV-FIMP DM, which naturally occurs when $m_{\Phi}\ll {\cal O}(\rm TeV)$~\footnote{A previous paper [24] investigated the case of an IR-FIMP DM $N_i$ in this model, simply by setting the Yukawa couplings $Y^N_{\alpha_i}\ll 1$. See also some related works [18].}. Before entering the details, we would like to add additional theoretical motivations for the decoupling scenario. Actually, there is no convincing reason that should lie around the weak scale. On the contrary, if we push much above the TeV scale, light active neutrino mass seems more natural since the model somehow becomes the radiative seesaw. Moreover, it greatly suppresses the LFV rates and avoids the potential parity-violating issue~\cite{Lindner:2016kqk}. Now we analyze if the UV-FIMP is able to gain correct relic density in the usual UV-FIMP scenario where the DM mass lies much below the reheating temperature. The answer is marginally positive. To see this, let us start from the dimension-six operators which account for DM UV freeze-in production,
\begin{align}\label{}
\f{Y_{\alpha_i}^NY_{\beta_j}^{N*}}{m_{\Phi}^2}\L\bar\ell_\alpha P_R N_i\R \L\bar  N_j P_L \ell_\beta \R\equiv
 \f{1}{\Ld_{\alpha\beta,ij}^2} \L\bar\ell_\alpha P_R N_i\R \L\bar  N_j  P_L\ell_\beta \R,
\end{align}
obtained after integrating out the heavy. Sterile neutrinos are produced via the scattering processes $\ell_\alpha\bar\ell_\beta \ra N_iN_j$. Note that these operators also furnish the decay channels for the 
heavier sterile neutrino states: $N_{2, 3}\ra N_1 +\bar\ell_\alpha+ \ell_\beta$. Typically, $N_{2, 3}$ are supposed to be similar to $N_1$ (the lightest sterile neutrino, the DM candidate) and they are also frozen-in and then transfer their number densities to $N_1$. Therefore we should summer over all three flavors. In the high reheating temperature scenario $T_{RH}\gg M_N$, at very high energy one can treat all particles massless in the freeze-in processes. Applying the general result Eq.~(\ref{d5:omega}) to this case, one gets the relic density 
\begin{align}\label{}
\Omega_{N_1}h^2=g_Ag_B\f{g_*(T_0)}{g_*(T_{RH})}\f{T_0^3T_{RH}^3}{C_r\rho_c}\f{3!2!}{3\times 2^3\pi^5}
\sum\f{M_1}{\Ld_{\alpha\beta,ij}^4}h^2\equiv F \sum\f{M_1}{\Ld_{\alpha\beta,ij}^4}.
\end{align}
It is larger than the contribution from certain flavors $\alpha, \beta$ and $i=i=k$. On the other
hand, barring fine-tuning of couplings, we may estimate the magnitude of order of active neutrino mass Eq. (\ref{nu:mass}). Without loss of generality, we assume mass order $M_3 > M_2 > M_1$, and then one can derive the inequality:
\begin{align}\label{}
(M_\nu)_{\alpha\beta}>&\f{\ld_5v^2}{16\pi^2}\sqrt{M_1}\sum_k\f{\sqrt{M_1}}{\Ld_{\alpha\beta,ij}^2}>
\f{\ld_5v^2}{16\pi^2}\sqrt{M_1}\sqrt{ \f{\Omega_{N_1}h^2}{F}}\cr
\sim&0.6\times10^{-4}\L\f{\ld_5}{1}\R \L\f{M_1}{0.1\rm TeV}\R^{\f{1}{2}} \L\f{\Omega_{N_1}h^2}{0.1}\R^{\f{1}{2}} 
 \L\f{1\rm TeV}{T_{RH}}\R^{\f{3}{2}}  \L\f{g_{*}(T_{RH})}{106}\R^{\f{3}{4}}\rm eV, 
\end{align}
where we have taken $g_A = g_B = 2$ for the left-handed lepton and as a factor 2 for the doublet.

The above inequality indicates that the correct relic density of $N_1$ is marginally consistent with the heaviest neutrino mass scale, $\sim 10^{-2}$ eV, if the reheating temperature is around the TeV scale; moreover, a mass hierarchy among $N_i$ is strongly favored, e.g., $N_3$ is at least two orders of magnitude heavier than $N_1$ thus in favor of enhancing the neutrino mass scale. Note that the effective suppression scale is 
\begin{align}\label{}
\L \Ld_{\alpha\beta,ij}^4 \R^{1/4}=\L\f{FM_1}{\Omega_{N_1}h^2}\R^{1/4}=1.8\times10^9\rm GeV,
\end{align}
where we have used the same parameterization in Eq. (3.9). It is indeed much higher than $T_{RH}$. A satisfying parameter setup is as: $m_\Phi \sim {\cal O} (10^6) $ GeV and $Y^N\sim{\cal O}(0.1)$, which could easily evade the LFV bound and moreover will not give rise to a very serious new  fine-tuning source for the SM Higgs. 

\subsubsection{Other options from production during reheating}

DM can be produced by other ways in this model. The first possibility is by means of the decay of the heavy mediator, which can be abundantly produced during reheating. Actually, for the above scenario, we have to check if the mediator is sufficiently heavy that its contribution to DM relic density is negligible. This could be done through Eq. (3.11), 
\begin{align}\label{}
\Omega_{N_1}h^2 \sim 3\times10^{-4} \L\f{M_1}{0.1\rm TeV}\R\L\f{10^3T_{RH}}{m_\Phi}\R^7 \L\f{10^6\rm GeV}{m_\Phi}\R
\L\f{\alpha_2}{0.01}\R\L\f{106}{g} \R^{3/2},
\end{align}
which is indeed a tiny contribution. We explain some details about the above estimate. $\Phi$ is participating in weak gauge interactions, and thus the radiation background components, the weak gauge bosons $V$ and SM fermions could produce via scattering with cross sections $\sim (g_2^4/16\pi)/m_\Phi^2$. Nevertheless, it is seen that the above contribution can be easily enhanced by considering a lighter $m_\Phi$, says a few times lighter than $10^3T_{RH}$. As a matter of fact, the light $N_1$ scenario tends to overproduce DM instead of the other way around. So, this way may be more interesting when $N_1$ is heavier than $T_{RH}$ and as a result the yield during
the radiation dominating era is highly suppressed; we have to fall back on DM production during reheating, either indirectly (namely the mediator decay) or directly, discussed in the following.

As stated before, in this model the sterile neutrino DM being a conventional UV-FIMP candidate tends to overclose the Universe. It drives us to consider the possibility that $N_1$  is heavier rather than lighter compared to the reheating temperature, and consequently its production during the radiation dominating era becomes negligible; moreover, the mediator mass is still very heavy and hence the above contribution is also negligible. We then fall back on the direct UV freeze-in of $N_1$  during the reheating era. According to Eq. (\ref{d5:omega}), the relic density is estimated to be
\begin{equation}
\Omega_{X}h^{2}\approx 2.3\times \L\frac{106}{g_{*}}\R^{3/2}\L\frac{5T_{RH}}{M_{1}} \R^{3}\L\frac{10^{5}T_{RH}}{\Lambda_{eff}}\R^{4}\,.
\end{equation}
Taking typical values for  $T_{RH}\gtrsim 1$ TeV (It can be even lower, but that may render the electroweak phase transition problematic.), $m_{\Phi}\sim 10^{6}$ GeV and $Y_{N}\sim 0.1$, it is checked 
that correct DM relic density can be achieved. The resulting neutrino mass in this scenario tends to be larger, 
but it can be easily made smaller via a smaller $\lambda_{5}$.

\section{Conclusions}

The FIMP is a competing candidate to dark matter, and it even becomes more and more attractive compared to the conventional WIMP DM, which has been seriously challenged by more and more null DM detection experiments; FIMP DM by definition barely leaves detectable signatures at the conventional detectors. The most popular FIMP DM is based on the renormalizable operators which freeze-in DM dominantly in the IR region, insensitive to the ultimately hot early Universe. However, usually it suffers the drawback of needing tiny couplings. This can be overcame in the UV-FIMP scenario based on high-dimensional operators, which, due to the strong dependence on temperature of amplitudes, instead freeze-in DM at UV.  The previous works impose the UV cut-off at the reheating temperature $T_{RH}$. Motivated by its UV-sensitivity, we investigate the effects from the even earlier Universe, i.e., the reheating era, and find that:
\begin{itemize}
\item In the usual case with $T_{RH}\gg m_{\rm DM}$, the production rate during reheating is indeed negligible as long as $d\leq 8$. 
\item The mediator, which may be sufficiently light in the concrete model and then can be abundantly produced during reheating. 
\item In particular, when $T_{RH}$ is even lower than $m_{\rm DM}$ and DM can be directly produced during reheating if its mass does not exceed $T_{MAX}$.  
\end{itemize}
We apply the general discussions in two concrete models which respectively are based on the $d=5$ and $d=6$ operators, both connected to neutrino mass origins.

\vskip 20pt

\noindent {\bf{Note added}} During the completion of this paper, we noticed that two recent papers~\cite{Kolb:2017jvz, Tang:2017hvq} studied nonthermal dark matter thermal production through higher dimensional Higgs portal operators suppressed by Planck scale, which actually is a special example of our study. In particular,  Ref.~\cite{Kolb:2017jvz}  also made a detailed analysis on superheavy DM production during reheating.

\vskip 18pt

\noindent {\bf{Acknowledgements}}
This work is supported in part by the National Science Foundation of China (11775086, 11775093, 11422545). ZK would like to greatly thank KIAS for the hospitality during his visiting, when the paper is finalized. 

\appendix

\section{Operators and reaction rates}

In the low reheating scenarios DM mass should be taking into account, which may lead to modifications to the simple amplitude squared ansatz adopted in the context. In this appendix we show the form of modifications based on several typical operators which appear in the literatures frequently. It is difficult to exhaust all kinds of operators. For a specific concrete model a detailed calculation is needed. For demonstration, we list some examples below:
\begin{table}[ht]
\begin{center}
\begin{tabular}{l|l}
${\cal O}_{1}=\bar f f \bar XX$ \quad&\quad $s(s-4M_{X}^{2})$\cr
${\cal O}_{2}=\bar f\gamma_{5} f \bar X\gamma_{5}X$ \quad&\quad $s^{2}$\cr
${\cal O}_{3}=\bar f \gamma^{\mu}f \bar X\gamma_{\mu}X$ \quad&\quad $s(s+4M_{X}^{2}+\frac{s-4M_{X}^{2}}{3})$\cr
${\cal O}_{4}=\bar f \gamma^{\mu}\gamma_{5}f \bar X\gamma_{\mu}\gamma_{5}X$ \quad&\quad $s(s+4M_{X}^{2}+\frac{s-4M_{X}^{2}}{3})$\cr
${\cal O}_{5}=|H|^{2} \bar XX$ \quad&\quad $s-4M_{X}^{2}$\cr
${\cal O}_{6}=H^{\dag}\overleftrightarrow{\partial_{\mu}} H  \bar X\gamma^{\mu}X$ \quad&\quad $s(s+2M_{X}^{2})$\cr
${\cal O}_{7}=H^{\dag}\overleftrightarrow{\partial_{\mu}} H  \bar X\gamma^{\mu}\gamma_{5}X$ \quad&\quad $s(s-4M_{X}^{2})$
\end{tabular}
\end{center}
\end{table}%
We only show the $s-$dependent parts of the amplitude squared, which are important in calculating the reaction density. It can be seen that the deviation to the naive power ansatz $s^{d-4}$ arises due to the heavy fermionic DM mass, in the form of a polynomial of $M_{X}^{2}/s$, which would make numerically difference near the mass threshold. In particular, it may be a suppression factor, namely DM velocity square $(1-4M_{X}^{2}/s)$. Therefore,
the resulting reaction density in general may involve an integral with a form
\begin{align}
\widetilde{\gamma}&=\int^{\infty}_{2M_{X}} ds s^{n/2} (1-4M_{X}^{2}/s)^{m/2} K_{1}(\sqrt{s}/T)\cr
&=T^{n+2}2^{m-2}\Gamma(1+\frac{m}{2})\times \cr
& z^{1+n}\text{MeijerG}[\{\{\},\{\frac{1}{2}(1+m-n)\}\} ,\{\{ \frac{-1+m}{2},\frac{1+m}{2},-\frac{1+n}{2}  \},\{ \} \},\frac{z^{2}}{4}]\,,
\end{align}
with $m=1$ or $3$, $n$ belongs to integer and $z=2M_{X}/T$; it is denoted as $M(m, n; z)$. The MeijerG function further admits an asymptotic expansion which works for all $z$, for instance
\begin{align}
M(3, 1; z)\approx & \frac{\sqrt{\pi } e^{-z} \left(\frac{165}{z^3}+\frac{162}{z^2}+\frac{72}{z}+16\right)}{8 z}, \cr
M(3, 3; z)\approx &\frac{1}{128} \sqrt{\pi } e^{-z} \left(\frac{10395}{z^4}+\frac{10440}{z^3}+\frac{4560}{z^2}+\frac{1088}{z}+128\right), \\
M(5, 1; z)\approx & \frac{\sqrt{\pi } e^{-z}
   \left(\frac{62685}{z^5}+\frac{62550}{z^4}+\frac{30000}{z^3}+\frac{9120}{z^2}+\frac{1920}{z}+256\right)}{128 z}, \cr
M(5, 3; z)\approx & \frac{\sqrt{\pi } e^{-z}
   \left(\frac{4003965}{z^6}+\frac{4005540}{z^5}+\frac{1877400}{z^4}+\frac{542400}{z^3}+\frac{105600}{z^2}+\frac{13824}{z}+10
   24\right)}{1024}, 
\end{align}
Note that for $m=1$ and $3$ the expansion (in the brackets) terminates at ${\cal O}(z^{-n})$ and ${\cal O}(z^{-n-1})$, respectively. One can use the above expansions to calculate the final integral over $T$ (or $z$). Or one can directly
implement the integral over $z$ in terms of the complete MeijerG function (times a factor $z^k$ with $k$ a positive integer), to obtain the primitive function
\begin{align}
\f{z^{1+k}}{2}\text{MeijerG}[\{\{\f{1-k}{2} \},\{\frac{1}{2}(1+m-n)\}\} ,\{\{ \frac{-1+m}{2},\frac{1+m}{2},-\frac{1+n}{2}  \},
\{-\f{1+k}{2} \} \},\frac{z^{2}}{4}].\,
\end{align}


\vspace{-.3cm}
\begin{thebibliography}{99}


%%%%%%%%%%%%%%%%%%%%%%%%%%%%%%%%%%%%%%% ÃÃœÃÃÃ 

\bibitem{Arcadi:2017kky} 
  G.~Arcadi, M.~Dutra, P.~Ghosh, M.~Lindner, Y.~Mambrini, M.~Pierre, S.~Profumo and F.~S.~Queiroz,
  %``The Waning of the WIMP? A Review of Models, Searches, and Constraints,''
  arXiv:1703.07364 [hep-ph].


%\cite{McDonald:2001vt}
\bibitem{McDonald:2001vt} 
  J.~McDonald,
  %``Thermally generated gauge singlet scalars as selfinteracting dark matter,''
  Phys.\ Rev.\ Lett.\  {\bf 88}, 091304 (2002)
  [hep-ph/0106249].

  
%\cite{Hall:2009bx}
\bibitem{Hall:2009bx} 
  L.~J.~Hall, K.~Jedamzik, J.~March-Russell and S.~M.~West,
  %``Freeze-In Production of FIMP Dark Matter,''
  JHEP {\bf 1003}, 080 (2010)
  [arXiv:0911.1120 [hep-ph]].
 
\bibitem{Bernal:2017kxu}
  N.~Bernal, M.~Heikinheimo, T.~Tenkanen, K.~Tuominen and V.~Vaskonen,
  %``The Dawn of FIMP Dark Matter: A Review of Models and Constraints,''
  arXiv:1706.07442 [hep-ph].

\bibitem{Klasen:2013ypa} 
  M.~Klasen and C.~E.~Yaguna,
  %``Warm and cold fermionic dark matter via freeze-in,''
  JCAP {\bf 1311}, 039 (2013). 

\bibitem{Kang:2014cia}
  Z.~Kang,
  %``Upgrading sterile neutrino dark matter to FI$m$P using scale invariance,''
  Eur.\ Phys.\ J.\ C {\bf 75}, no. 10, 471 (2015); %\cite{Kang:2015a\bibitem{Kang:2015aqa} 
%  Z.~Kang,
  %``View FImP miracle (by scale invariance) Ã  la self-interaction,''
  Phys.\ Lett.\ B {\bf 751}, 201 (2015). %[arXiv:1505.06554 [hep-ph]].

%\cite{Queiroz:2014yna}
\bibitem{Queiroz:2014yna} 
  F.~S.~Queiroz and K.~Sinha,
  %``The Poker Face of the Majoron Dark Matter Model: LUX to keV Line,''
  Phys.\ Lett.\ B {\bf 735}, 69 (2014).
  
  
  \bibitem{Bae:2017tqn} 
  K.~J.~Bae, A.~Kamada, S.~P.~Liew and K.~Yanagi,
  %``Colder Freeze-in Axinos Decaying into Photons,''
  arXiv:1707.02077 [hep-ph].

  
  \bibitem{Essig:2011nj} 
  R.~Essig, J.~Mardon and T.~Volansky,
  %``Direct Detection of Sub-GeV Dark Matter,''
  Phys.\ Rev.\ D {\bf 85}, 076007 (2012)
  [arXiv:1108.5383 [hep-ph]].
  
    \bibitem{Dev:2013yza}
  P.~S.~Bhupal Dev, A.~Mazumdar and S.~Qutub,
  %``Constraining Non-thermal and Thermal properties of Dark Matter,''
  Front.\ in Phys.\  {\bf 2}, 26 (2014)

\bibitem{Merle:2015oja} 
  A.~Merle and M.~Totzauer,
  %``keV Sterile Neutrino Dark Matter from Singlet Scalar Decays: Basic Concepts and Subtle Features,''
  JCAP {\bf 1506}, 011 (2015);  J.~König, A.~Merle and M.~Totzauer,
  %``keV Sterile Neutrino Dark Matter from Singlet Scalar Decays: The Most General Case,''
  JCAP {\bf 1611}, no. 11, 038 (2016).

\bibitem{Drewes:2015eoa} 
  M.~Drewes and J.~U.~Kang,
  %``Sterile neutrino Dark Matter production from scalar decay in a thermal bath,''
  JHEP {\bf 1605}, 051 (2016).
  
  
  \bibitem{Heeck:2017xbu} 
  J.~Heeck and D.~Teresi,
  %``Cold keV dark matter from decays and scatterings,''
  Phys.\ Rev.\ D {\bf 96}, no. 3, 035018 (2017).


\bibitem{Tenkanen:2016twd}
  T.~Tenkanen,
  %``Feebly Interacting Dark Matter Particle as the Inflaton,''
  JHEP {\bf 1609}, 049 (2016)


\bibitem{Kainulainen:2016vzv}
  K.~Kainulainen, S.~Nurmi, T.~Tenkanen, K.~Tuominen and V.~Vaskonen,
  %``Isocurvature Constraints on Portal Couplings,''
 JCAP {\bf 1606}, no. 06, 022 (2016)

\bibitem{Roland:2016gli}
  S.~B.~Roland and B.~Shakya,
  %``Cosmological Imprints of Frozen-In Light Sterile Neutrinos,''
  JCAP {\bf 1705}, no. 05, 027 (2017).

\bibitem{Chu1}
  X.~Chu, T.~Hambye and M.~H.~G.~Tytgat,
  %``The Four Basic Ways of Creating Dark Matter Through a Portal,''
  JCAP {\bf 1205}, 034 (2012)
  [arXiv:1112.0493 [hep-ph]].
  
  
  
  \bibitem{Kaneta:2016vkq}
  K.~Kaneta, Z.~Kang and H.~S.~Lee,
  %``Right-handed neutrino dark matter under the $B ? L$ gauge interaction,''
  JHEP {\bf 1702}, 031 (2017).

\bibitem{Ayazi:2015jij}
  S.~Yaser Ayazi, S.~M.~Firouzabadi and S.~P.~Zakeri,
  %``Freeze-in production of Fermionic Dark Matter with Pseudo-scalar and Phenomenological Aspects,''
  J.\ Phys.\ G {\bf 43}, no. 9, 095006 (2016).

\bibitem{Hardy:2017wkr} 
  E.~Hardy and J.~Unwin,
  %``Symmetric and Asymmetric Reheating,''
  JHEP {\bf 1709}, 113 (2017).


 \bibitem{Duch:2017khv} 
  M.~Duch, B.~Grzadkowski and D.~Huang,
  %``Strongly self-interacting vector dark matter via freeze-in,''
  arXiv:1710.00320 [hep-ph].

 
  %\cite{Chung:1998rq}
\bibitem{Chung:1998rq} 
  D.~J.~H.~Chung, E.~W.~Kolb and A.~Riotto,
  %``Production of massive particles during reheating,''
  Phys.\ Rev.\ D {\bf 60}, 063504 (1999)
  [hep-ph/9809453].
  
  \bibitem{Roland:2014vba} 
  S.~B.~Roland, B.~Shakya and J.~D.~Wells,
  %``Neutrino Masses and Sterile Neutrino Dark Matter from the PeV Scale,''
  Phys.\ Rev.\ D {\bf 92}, no. 11, 113009 (2015)
  [arXiv:1412.4791 [hep-ph]].


\bibitem{Ellis:1984eq}
  J.~R.~Ellis, J.~E.~Kim and D.~V.~Nanopoulos,
  %``Cosmological Gravitino Regeneration and Decay,''
  Phys.\ Lett.\  {\bf 145B}, 181 (1984).

\bibitem{inPQ}
E. J. Chun, H. B. Kim and D. H. Lyth, Phys. Rev. D 62 (2000) 125001;
% [hep-ph/0008139].
L. Covi, H. B. Kim, J. E. Kim and L. Roszkowski,
%Axinos as dark matter,
JHEP 0105 (2001) 033; %[hep-ph/0101009].
K. J. Bae, K. Choi and S. H. Im, %Eective Interactions of Axion Supermultiplet and Thermal Production of Axino Dark Matter,
JHEP 1108 (2011) 065. % [1106.2452].

\bibitem{Elahi:2014fsa}
  F.~Elahi, C.~Kolda and J.~Unwin,
  %``UltraViolet Freeze-in,''
  JHEP {\bf 1503}, 048 (2015).

 \bibitem{McDonald:2015ljz}
  J.~McDonald,
  %``Warm Dark Matter via Ultra-Violet Freeze-In: Reheating Temperature and Non-Thermal Distribution for Fermionic Higgs Portal Dark Matter,''
  JCAP {\bf 1608}, no. 08, 035 (2016).



\bibitem{Co:2015pka} 
  R.~T.~Co, F.~D'Eramo, L.~J.~Hall and D.~Pappadopulo,
  %``Freeze-In Dark Matter with Displaced Signatures at Colliders,''
  JCAP {\bf 1512}, no. 12, 024 (2015).

\bibitem{Garcia:2017tuj} 
  M.~A.~G.~Garcia, Y.~Mambrini, K.~A.~Olive and M.~Peloso,
  %``Enhancement of the Dark Matter Abundance Before Reheating: Applications to Gravitino Dark Matter,''
  arXiv:1709.01549 [hep-ph].


\bibitem{Brdar:2017wgy} 
  V.~Brdar, J.~Kopp, J.~Liu and X.~P.~Wang,
  %``Return of the X-rays: A New Hope for Fermionic Dark Matter at the keV Scale,''
  arXiv:1710.02146 [hep-ph].

  %\cite{Ma:2006km}
\bibitem{Ma:2006km} 
  E.~Ma,
  %``Verifiable radiative seesaw mechanism of neutrino mass and dark matter,''
  Phys.\ Rev.\ D {\bf 73}, 077301 (2006)
  [hep-ph/0601225].


\bibitem{Davidson:2008bu} 
  S.~Davidson, E.~Nardi and Y.~Nir,
  %``Leptogenesis,''
  Phys.\ Rept.\  {\bf 466}, 105 (2008)
  [arXiv:0802.2962 [hep-ph]].

\bibitem{Baek:2015fma} 
  S.~Baek and Z.~F.~Kang,
  %``Naturally Large Radiative Lepton Flavor Violating Higgs Decay Mediated by Lepton-flavored Dark Matter,''
  JHEP {\bf 1603}, 106 (2016)
  [arXiv:1510.00100 [hep-ph]].
  

 \bibitem{Okada:2015kkj}
  H.~Okada, Y.~Orikasa and T.~Toma,
  %``Nonthermal dark matter models and signals,''
  Phys.\ Rev.\ D {\bf 93}, no. 5, 055007 (2016).

\bibitem{Adam:2013mnn} 
  J.~Adam {\it et al.} [MEG Collaboration],
  %``New constraint on the existence of the $\mu^+ \to e^+\gamma$ decay,''
  Phys.\ Rev.\ Lett.\  {\bf 110}, 201801 (2013)
  [arXiv:1303.0754 [hep-ex]].

\bibitem{Kubo:2006yx} 
  J.~Kubo, E.~Ma and D.~Suematsu,
  %``Cold Dark Matter, Radiative Neutrino Mass, $\mu \to e\gamma$, and Neutrinoless Double Beta Decay,''
  Phys.\ Lett.\ B {\bf 642}, 18 (2006)
  [hep-ph/0604114].


\bibitem{Lindner:2016kqk} 
  M.~Lindner, M.~Platscher, C.~E.~Yaguna and A.~Merle,
  %``Fermionic WIMPs and vacuum stability in the scotogenic model,''
  Phys.\ Rev.\ D {\bf 94}, no. 11, 115027 (2016).



\bibitem{Kolb:2017jvz} 
  R.~W.~Kolb and A.~J.~Long,
  %``Superheavy dark matter through Higgs portal operators,''
  arXiv:1708.04293 [astro-ph.CO].
  
  \bibitem{Molinaro:2014lfa} 
  E.~Molinaro, C.~E.~Yaguna and O.~Zapata,
  %``FIMP realization of the scotogenic model,''
  JCAP {\bf 1407}, 015 (2014) 
  [arXiv:1405.1259 [hep-ph]].
  
  
\bibitem{FIMPdecay}
Z. Kang and T. Li, %Decaying Dark Matter in the Supersymmetric Standard Model with Freeze-in and Seesaw mechanims,
JHEP 02 (2011) 035.


\bibitem{Kolb:2017jvz} 
  R.~W.~Kolb and A.~J.~Long,
  %``Superheavy dark matter through Higgs portal operators,''
  arXiv:1708.04293 [astro-ph.CO].

\bibitem{Tang:2017hvq} 
  Y.~Tang and Y.~L.~Wu,
  %``On Thermal Gravitational Contribution to Particle Production and Dark Matter,''
  Phys.\ Lett.\ B {\bf 758}, 402 (2016);
  Phys.\ Lett.\ B {\bf 774}, 676 (2017).
 


 \end{thebibliography}
\end{document}